\documentclass[letterpaper,10pt]{IEEEtran}
\usepackage{amsmath,graphicx,epsfig,color,amsfonts}
\usepackage{version}
\def\prob{\mathbb{P}}

\newcommand{\Limits}[3]{\ensuremath{\underset{#1}{\overset{#2}{#3}}}}

\newcommand{\Exp}[1]{\ensuremath{\mathbb{E}\left[#1\right]}}

\def\dist{\mathcal{D}}

\def\prob{\mathbb{P}}
\def\real{\mathbb{R}}
\newcommand{\until}[1]{\{1,\dots, #1\}}
\newcommand{\untill}[1]{\{0,\dots, #1\}}

\newcommand{\setdef}[2]{\{#1 \; | \; #2\}}

\newcommand{\map}[3]{#1: #2 \rightarrow #3}

\newcommand{\subject}{\text{subject to}}

\newcommand{\minimize}{\text{minimize}}

\newcommand\oprocendsymbol{\hbox{$\square$}}
\newcommand\oprocend{\relax\ifmmode\else\unskip\hfill\fi\oprocendsymbol}

\renewcommand{\theenumi}{\roman{enumi}}

\newtheorem{lemma}{Lemma}
\newtheorem{theorem}{Theorem}

\title{Randomized Sensor Selection \\ in Sequential Hypothesis Testing}

\author{Vaibhav~Srivastava \hspace{1in} Kurt~Plarre \hspace{1in}
  Francesco~Bullo \thanks{This work has been supported in part by AFOSR
    MURI Award F49620-02-1-0325.}  \thanks{Vaibhav Srivastava and Francesco
    Bullo are with the Center for Control, Dynamical Systems, and
    Computation, University of California, Santa Barbara, Santa Barbara, CA
    93106, USA,\tt{ \{vaibhav,bullo\}@engineering.ucsb.edu}}\thanks{Kurt
    Plarre is with the Department of Computer Science, University of
    Memphis, Memphis, TN 38152, USA, \tt{kplarre@memphis.edu}} }

\begin{document}
\maketitle

\begin{abstract}
  We consider the problem of sensor selection for time-optimal detection of
  a hypothesis. We consider a group of sensors transmitting their
  observations to a fusion center. The fusion center considers the output
  of only one randomly chosen sensor at the time, and performs a sequential
  hypothesis test. We consider the class of sequential tests which are easy
  to implement, asymptotically optimal, and computationally amenable.  For
  three distinct performance metrics, we show that, for a generic set of
  sensors and binary hypothesis, the fusion center needs to consider at
  most two sensors. We also show that for the case of multiple hypothesis,
  the optimal policy needs at most as many sensors to be observed as the
  number of underlying hypotheses.
\end{abstract}


\section{INTRODUCTION}
In today's information-rich world, different sources are best informers
about different topics. If the topic under consideration is well known
beforehand, then one chooses the best source. Otherwise, it is not obvious
what source or how many sources one should observe. This need to identify
sensors (information sources) to be observed in decision making problems is
found in many common situations, e.g., when deciding which news channel to
follow. When a person decides what information source to follow, she relies
in general upon her experience, i.e., one knows through experience what
combination of news channels to follow.

In engineering applications, a reliable decision on the underlying
hypothesis is made through repeated measurements.  Given infinitely many
observations, decision making can be performed accurately. Given a cost
associated to each observation, a well-known tradeoff arises between
accuracy and number of iterations. Various sequential hypothesis tests have
been proposed to detect the underlying hypothesis within a given degree of
accuracy. There exist two different classes of sequential tests. The first
class includes sequential tests developed from the dynamic programming
point of view. These tests are optimal and, in general, difficult to
implement~\cite{CWB-VVV:94}. The second class consists of
easily-implementable and asymptotically-optimal sequential tests; a
widely-studied example is the Sequential Probability Ratio Test (SPRT) for
binary hypothesis testing and its extension, the Multi-hypothesis
Sequential Probability Ratio Test (MSPRT).

In this paper, we consider the problem of quickest decision making using
sequential probability ratio tests. Recent advances in cognitive
psychology~\cite{RB-EB-etal:06} show that the performance of a human
performing decision making tasks, such as "two-alternative forced choice
tasks," is well modeled by a drift diffusion process, i.e., the
continuous-time version of SPRT.  Roughly speaking, modeling decision
making as an SPRT process is somehow appropriate even for situations in
which a human is making the decision. 


Sequential hypothesis testing and quickest detection problems have been
vastly studied~\cite{HVP-OH:08,MB-IVN:93}. The SPRT for binary decision
making was introduced by Wald in \cite{AW:45}, and was extended by Armitage
to multiple hypothesis testing in~\cite{PA:50}. The Armitage test, unlike
the SPRT, is not necessarily optimal~\cite{XJZ:89}. Various other tests for
multiple hypothesis testing have been developed throughout the years; a
survey is presented in~\cite{JPS:95}. Designing hypothesis tests, i.e.,
choosing thresholds to decide within a given expected number of iterations,
through any of the procedures in~\cite{JPS:95} is infeasible as none of
them provides any results on the expected sample size.  A sequential test
for multiple hypothesis testing was developed
in~\cite{CWB-VVV:94},~\cite{VPD-AGT-VVV:99},~and~\cite{VPD-AGT-VVV:00},
which provides with an asymptotic expression for the expected sample
size. This sequential test is called the MSPRT and reduces to the SPRT in
case of binary hypothesis.

Recent years have witnessed a significant interest in the problem of sensor
selection for optimal detection and estimation. Tay et
al~\cite{WPT-JNT-MZW:07} discuss the problem of censoring sensors for
decentralized binary detection. They assess the quality of sensor data by
the Neyman-Pearson and a Bayesian binary hypothesis test and decide on
which sensors should transmit their observation at that time instant. Gupta
et al~\cite{VG-THC-BH-RMM:06} focus on stochastic sensor selection and
minimize the error covariance of a process estimation problem. Isler et
al~\cite{VI-RB:06} propose geometric sensor selection schemes for error
minimization in target detection. Debouk et al~\cite{RD-SL-DT:02} formulate
a Markovian decision problem to ascertain some property in a dynamical
system, and choose sensors to minimize the associated cost.  Wang et
al~\cite{HW-KY-GP-DE:04} design entropy-based sensor selection algorithms
for target localization.  Joshi et al~\cite{SJ-SB:09} present a convex
optimization-based heuristic to select multiple sensors for optimal
parameter estimation. Bajovi\'c et al~\cite{DB-BS-JX:09} discuss sensor
selection problems for Neyman-Pearson binary hypothesis testing in wireless
sensor networks.

A third and last set of references related to this paper are those on
linear-fractional programming.  Various iterative and cumbersome algorithms
have been proposed to optimize linear-fractional functions \cite{SB-LV:04},
\cite{EB:03}. In particular, for the problem of minimizing the sum and the
maximum of linear-fractional functionals, some efficient iterative
algorithms have been proposed, including the algorithms by Falk et
al~\cite{JEF-SWP:92} and by Benson~\cite{HPB:04}.

In this paper, we analyze the problem of time-optimal sequential decision
making in the presence of multiple switching sensors and determine a sensor
selection strategy to achieve the same. We consider a sensor network where
all sensors are connected to a fusion center. The fusion center, at each
instant, receives information from only one sensor. Such a situation arises
when we have interfering sensors (e.g., sonar sensors), a fusion center
with limited attention or information processing capabilities, or sensors
with shared communication resources.  The fusion center implements a
sequential hypothesis test with the gathered information.  We consider two
such tests, namely, the SPRT and the MSPRT for binary and multiple
hypothesis, respectively. First, we develop a version of the SPRT and the
MSPRT where the sensor is randomly switched at each iteration, and
determine the expected time that these tests require to obtain a decision
within a given degree of accuracy. Second, we identify the set of sensors
that minimize the expected decision time. We consider three different cost
functions, namely, the conditioned decision time, the worst case decision
time, and the average decision time. We show that the expected decision
time, conditioned on a given hypothesis, using these sequential tests is a
linear-fractional function defined on the probability simplex. We exploit
the special structure of our domain (probability simplex), and the fact
that our data is positive to tackle the problem of the sum and the maximum
of linear-fractional functionals analytically. Our approach provides
insights into the behavior of these functions. The major contributions of
this paper are:
\begin{enumerate}
\item We develop a version of the SPRT and the MSPRT where the sensor is
  selected randomly at each observation.
\item We determine the asymptotic expressions for the thresholds and the
  expected sample size for these sequential tests.
\item We incorporate the processing time of the sensors into these models
  to determine the expected decision time.
\item We show that, to minimize the conditioned expected decision time, the
  optimal policy requires only one sensor to be observed.
\item We show that, for a generic set of sensors and $M$ underlying
  hypotheses, the optimal average decision time policy requires the fusion
  center to consider at most $M$ sensors. 


\item For the binary hypothesis case, we identify the optimal set of
  sensors in the worst case and the average decision time minimization
  problems. Moreover, we determine an optimal probability distribution for
  the sensor selection.

\item In the worst case and the average decision time minimization
  problems, we encounter the problem of minimization of sum and maximum of
  linear-fractional functionals. We treat these problems analytically, and
  provide insight into their optimal solutions.
\end{enumerate}

The remainder of the paper is organized in following way. In
Section~\ref{sec:problem-setup}, we present the problem setup. Some
preliminaries are presented in Section~\ref{sec:preliminaries}. We develop
the switching-sensor version of the SPRT and the MSPRT procedures in
Section~\ref{sec:switch-sensors}. In
Section~\ref{sec:optimal-sensor-selection}, we formulate the optimization
problems for time-optimal sensor selection, and determine their
solution. We elucidate the results obtained through numerical examples in
Section~\ref{sec:numerical-examples}. Our concluding remarks are in
Section~\ref{sec:conclusions}.

\section{PROBLEM SETUP}
\label{sec:problem-setup}
We consider a group of $n>1$ agents (e.g., robots, sensors, or cameras),
which take measurements and transmit them to a fusion center. We generically call these agents "sensors." We identify the fusion center with a person supervising the agents, and call it "the supervisor."

\begin{figure}[ht]
  \begin{center}
    \scalebox{0.25}{\includegraphics{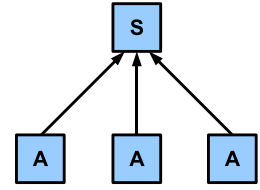}}
  \end{center}
  \caption{The agents $A$ transmit their observation to the supervisor $S$, one at the time. The supervisor performs a sequential hypothesis test to decide on the underlying hypothesis.} \label{figure}
\end{figure}

The goal of the supervisor is to decide, based on the measurements it
receives, which of the $M\ge 2$ alternative hypotheses or ``states of
nature,'' $H_k$, $k\in\untill{M-1}$ is correct. For doing so, the
supervisor uses sequential hypothesis tests, which we briefly review in the
next section.

We assume that only one sensor can transmit to the supervisor at each (discrete)
time instant. Equivalently, the supervisor can process data from only one of the
$n$ agents at each time. Thus, at each time, the supervisor must decide which
sensor should transmit its measurement. We are interested in finding the
optimal sensor(s), which must be observed in order to minimize the decision
time.

We model the setup in the following way:
\begin{enumerate}
\item Let $s_l \in \until{n}$ indicate which sensor transmits its
  measurement at time instant $l\in\mathbb{N}$.

\item Conditioned on the hypothesis $H_k$, $k\in\{0,\ldots,M-1\}$, the
  probability that the measurement at sensor $s$ is $y$, is denoted by
  $f_s^k(y)$.

\item The prior probability of the hypothesis $H_k$, $k\in\untill{M-1}$,
  being correct is $\pi_k$.

\item The measurement of sensor $s$ at time $l$ is $y_{s_l}$. We assume
  that, conditioned on hypothesis $H_k$, $y_{s_l}$ is independent of
  $y(s_{\bar{l}})$, for $(l,s_l) \neq (\bar{l},s_{\bar{l}})$.

\item The time it takes for sensor $s$ to transmit its measurement (or for
  the supervisor to process it) is $T_s>0$.

\item The supervisor chooses a sensor randomly at each time instant; the
  probability to choose sensor $s$ is stationary and given by $q_s$.

\item The supervisor uses the data collected to execute a sequential
  hypothesis test with the desired probability of incorrect decision,
  conditioned on hypothesis $H_k$, given by $\alpha_k$.

\item We assume that there are no two sensors with identical conditioned
  probability distribution $f_s^k(y)$ and processing time $T_s$. If there
  are such sensors, we club them together in a single node, and distribute
  the probability assigned to that node equally among them.
\end{enumerate}

\section{PRELIMINARIES}
\label{sec:preliminaries}
\subsection{Linear-fractional function }
Given parameters $A\in\mathbb{R}^{q\times p}$, $B\in\mathbb{R}^q$,
$c\in\mathbb{R}^p$, and $d\in \mathbb{R}$, the function $\map{g}{
  \setdef{z\in\mathbb{R}^p}{c^Tz+d>0} } {\mathbb{R}^q}$, defined by
\begin{equation*}
  g(x)=\frac{Ax+B}{c^Tx+d},
\end{equation*}
is called a \emph{linear-fractional function} \cite{SB-LV:04}.  A
linear-fractional function is quasi-convex as well as quasi-concave. In
particular, if $q=1$, then any scalar linear-fractional function $g$ satisfies
\begin{equation}
  \label{eq:lin_frac}
  \begin{split}
    g(t x+(1-t) y) &\le \max\{g(x),g(y)\},\\
    g(t x+(1-t) y) &\ge \min\{g(x),g(y)\},
  \end{split}
\end{equation}
for all $t\in[0,1]$ and $x,y\in \setdef{z\in\mathbb{R}^p}{c^Tz+d>0}$.

\subsection{Kullback-Leibler distance}\label{subsec:kullback}
Given two probability distributions functions $\map{f_1}{\real}{\real_{\ge
    0}}$ and $\map{f_2}{\real}{\real_{\ge 0}}$, the Kullback-Leibler
distance $\map{\dist}{\mathcal{L}^1\times\mathcal{L}^1}{\real}$ is defined
by
\begin{equation*}
\dist(f_1,f_2)= \Exp{\log\frac{f_1(X)}{f_2(X)}}= \int_{\real} f_1(x) \log\frac{f_1(x)}{f_2(x)} dx.
\end{equation*}
Further, $\dist(f_1,f_2)\ge 0$, and the equality holds if and only if $f_1=f_2$ almost everywhere.

\subsection{Sequential Probability Ratio Test}
The SPRT is a sequential binary hypothesis test that provides us with two
thresholds to decide on some hypothesis, opposed to classical hypothesis
tests, where we have a single threshold. Consider two hypothesis $H_0$ and
$H_1$, with prior probabilities $\pi_0$ and $\pi_1$, respectively. Given
their conditional probability distribution functions $f(y|H_0)=f^0(y)$ and
$f(y|H_1)=f^1(y)$, and repeated measurements $\{y_1,y_2,\ldots \}$, with
$\lambda_0$ defined by
\begin{equation}\label{eq:prior}
  \lambda_0=\log\frac{\pi_1}{\pi_0},
\end{equation}
the SPRT provides us with two constants $\eta_0$ and $\eta_1$ to decide on
a hypothesis at each time instant $l$, in the following way:
\begin{enumerate}
\item Compute the log likelihood ratio: $\lambda_l$ := log $\frac{f^1(y_l)}{f^0(y_l)}$,
\item Integrate evidence up to time $N$, i.e., $\Lambda_N$ := $\Limits{l=0}{N} \sum \lambda_l$,
\item Decide only if a threshold is crossed, i.e., 
\begin{align*}
\begin{cases}
\Lambda_N > \eta_1 ,& \mbox{say } H_1, \\
\Lambda_N < \eta_0 , &  \mbox{say } H_0,\\
\Lambda_N \in {]\eta_0, \eta_1 [},&  \mbox{continue sampling}.
\end{cases}
\end{align*}
\end{enumerate}

Given the probability of false alarm $\prob(H_1|H_0)$ = $\alpha_0$ and
probability of missed detection $\prob(H_0|H_1)$ = $\alpha_1$, the Wald's
thresholds $\eta_0$ and $\eta_1$ are defined by
\begin{gather}\label{eq:thresholds}
  \eta_0=\log\frac{\alpha_1}{1-\alpha_0}, \quad \text{and} \quad \eta_1=\log\frac{1-\alpha_1}{\alpha_0}.
\end{gather}
The expected sample size $N$, for decision using SPRT is asymptotically
given by
\begin{equation}
  \label{eq:estimated_time}
  \begin{split}
    \Exp{N|H_0} &\to -\frac{(1-\alpha_0)\eta_0+ \alpha_0 \eta_1-\lambda_0}{\dist(f^0,f^1)},\quad \text{and}\\
    \Exp{N|H_1} &\to \frac{\alpha_1 \eta_0+(1-\alpha_1) \eta_1-\lambda_0}{\dist(f^1,f^0)},
  \end{split}
\end{equation}
as $-\eta_0,\eta_1\to\infty$.  The asymptotic expected sample size
expressions in equation~\eqref{eq:estimated_time} are valid for large
thresholds. The use of these asymptotic expressions as approximate expected
sample size is a standard approximation in the information theory
literature, and is known as Wald's
approximation~\cite{MB-IVN:93},~\cite{HVP-OH:08},~\cite{DS:85}.

For given error probabilities, the SPRT is the optimal sequential binary hypothesis test, if the sample size is considered as the cost function~\cite{DS:85}.

\subsection{Multi-hypothesis Sequential Probability Ratio Test}\label{subsec:msprt}
The MSPRT for multiple hypothesis testing was introduced
in~\cite{CWB-VVV:94}, and was further generalized in~\cite{VPD-AGT-VVV:99}
and~\cite{VPD-AGT-VVV:00}. It is described as follows.  Given $M$
hypotheses with their prior probabilities $\pi_k, k\in\untill{M-1}$, the
posterior probability after $N$ observations $y_l,l\in\until{N}$ is given
by
\begin{equation*}
  p_N^k=\prob(H=H_k|y_1,\ldots,y_N)=
  \frac{\pi_k\Limits{l=1}{N}{\Pi}f^k(y_l)}{\Limits{j=1}{M-1}{\sum}\pi_j\left(\Limits{l=1}{n}{\Pi}f^j(y_l)\right)}, 
\end{equation*}
where $f^k$ is the probability density function of the observation of the
sensor, conditioned on hypothesis $k$.

Before we state the MSPRT, for a given $N$, we define $\bar{k}$ by
\[
\bar{k}=\Limits{j\in\untill{M-1}}{}{\text{argmax }} \pi_j\Limits{l=1}{N}{\Pi} f^j(y_l).
\]

The MSPRT at each sampling iteration $l$ is defined as
\begin{align*}
\begin{cases}
  p^k_l > \frac{1}{1+\eta_k}, \text{for at least one } k ,& \text{say }H_{\bar{k}}, \\
  \text{otherwise,}& \mbox{continue sampling,}
\end{cases}
\end{align*}
where the thresholds $\eta_k$, for given frequentist error probabilities
(accept a given hypothesis wrongly) $\alpha_k$, $k\in\untill{M-1}$, are
given by
\begin{equation}\label{eq:threshold}
  \eta_k=\frac{\alpha_k}{\pi_k\gamma_k},
\end{equation}
where $\gamma_k\in{]0,1[}$ is a constant function of $f^k$
(see~\cite{CWB-VVV:94}).

It can be shown~\cite{CWB-VVV:94} that the expected sample size of the
MSPRT, conditioned on a hypothesis, satisfies
\begin{equation*}
  \Exp{N|H_k} \to \frac{-\log \eta_k}{\delta_k},
  \quad \text{as} \Limits{k\in{\untill{M-1}}}{}{\max} \eta_k \to 0^+,
\end{equation*}
where $\delta_k= \min\setdef{\dist(f^k,f^j)} {j\in\untill{M-1}\setminus
  \{k\} }$.

The MSPRT is an easily-implementable hypothesis test and is shown to be
asymptotically optimal in~\cite{CWB-VVV:94,VPD-AGT-VVV:99}.

\section{Sequential hypothesis tests with switching sensors}\label{sec:switch-sensors}
\subsection{SPRT with switching sensors}
Consider the case when the fusion center collects data from $n$ sensors. At each iteration the fusion center looks at one sensor chosen randomly with probability $q_s$, $s\in\until{n}$. The fusion center performs SPRT with the collected data. We define this procedure as SPRT with switching sensors. If we assume that  sensor $s_l$ is observed at iteration $l$, and the observed value is $y_{s_l}$, then SPRT with switching sensors is described as following, with the thresholds $\eta_0$ and $\eta_1$ defined in equation \eqref{eq:thresholds}, and $\lambda_0$ defined in equation~\eqref{eq:prior}:

\begin{enumerate}
\item Compute log likelihood ratio:
\[
\lambda_l := \log \frac{f_{s_l}^1(y_{s_l})}{f_{s_l}^0(y_{s_l})},
\]

\item Integrate evidence up to time $N$, i.e., $\Lambda_N$ := $\Limits{l=0}{N} \sum \lambda_l$,

\item Decide only if a threshold is crossed, i.e.,
\begin{align*}
\begin{cases}
\Lambda_N > \eta_1,& \mbox{say } H_1,\\
\Lambda_N < \eta_0,& \mbox{say } H_0,\\
\Lambda_N \in {]\eta_0, \eta_1 [},& \mbox{continue sampling}.
\end{cases}
\end{align*}
\end{enumerate}

\begin{lemma}[Expected sample size] 
  \label{lem:sample-size-switch-sprt} 
  For the SPRT with switching sensors described above, the expected sample
  size conditioned on a hypothesis is asymptotically given by:
  \begin{equation}
    \label{eq:iterations}
    \begin{split}
      \Exp{N|H_0} &\to -\frac{(1-\alpha_0)\eta_0+ \alpha_0
        \eta_1-\lambda_0}{\Limits{s=1}{n}{\sum}q_s
        \dist(f_s^0,f_s^1)},\quad \text{and}
      \\
      \Exp{N|H_1} &\to \frac{\alpha_1
        \eta_0+(1-\alpha_1)\eta_1-\lambda_0}{\Limits{s=1}{n}{\sum}q_s\dist(f_s^1,f_s^0)},
    \end{split}
  \end{equation}
  as $-\eta_0,\eta_1\to\infty$.
\end{lemma}
\begin{IEEEproof}
  Similar to the proof of Theorem 3.2 in ~\cite{MW:82}.
\end{IEEEproof}

The expected sample size converges to the values in
equation~\eqref{eq:iterations} for large thresholds. From
equation~\eqref{eq:thresholds}, it follows that large thresholds correspond
to small error probabilities.  In the remainder of the paper, we assume
that the error probabilities are chosen small enough, so that the above
asymptotic expression for sample size is close to the actual expected
sample size.

\begin{lemma}[Expected decision time] \label{lem:decision-time-switch-sprt}
Given the processing time of the sensors $T_s, s\in\until{n}$, the expected  decision time of the SPRT with switching sensors $T_d$, conditioned on the hypothesis $H_k$, $k\in\{0,1\}$, is 
\begin{equation}
  \label{eq:expected_decision_time}
  \Exp{T_d|H_k}=  \frac{\Limits{s=1}{n}\sum q_sT_s}{\Limits{s=1}{n}\sum
    q_sI^k_s}=\frac{q\cdot T}{q\cdot I^k},  \quad\mbox{for each }
  k\in\{0,1\}, 
\end{equation}
where $T,I^k\in\real^n_{>0}$, are constant vectors for each $k \in \{0,1\}$.
\end{lemma}
\begin{IEEEproof}
The decision time using SPRT with switching sensors is the sum of sensor's processing time at each iteration. We observe that the number of iterations in SPRT and the processing time of sensors are independent. Hence, the expected value of the decision time $T_d$ is
\begin{align}\label{eq:dec_time}
\Exp{T_d|H_k}= \Exp{N|H_k}\Exp{T}, \quad \text{for each } k\in\{0,1\}.
\end{align}
By the definition of expected value,
\begin{align}\label{eq:avg_proc_time}
\Exp{T} = \Limits{s=1}{n}\sum q_sT_s.
\end{align}
From equations \eqref{eq:iterations}, \eqref{eq:dec_time}, and    \eqref{eq:avg_proc_time} it follows that
\begin{equation*}
  \Exp{T_d|H_k}= \frac{\Limits{s=1}{n}{\sum}
    q_sT_s}{\Limits{s=1}{n}{\sum}q_sI^k_s}= \frac{q\cdot T}{q\cdot I^k},
  \quad\mbox{for each } k\in\{0,1\}, 
\end{equation*}
where $I^k_s\in\real_{>0}$ is a constant, for each $k \in \{0,1\}$, and
$s\in\until{n}$.
\end{IEEEproof}

\subsection{MSPRT with switching sensors}\label{subsec:dmsprt}

We call the MSPRT with the data collected from $n$ sensors while observing
only one sensor at a time as the MSPRT with switching sensors. The one
sensor to be observed at each time is determined through a randomized
policy, and the probability of choosing sensor $s$ is stationary and given
by $q_s$. Assume that the sensor $s_l\in\until{n}$ is chosen at time
instant $l$, and the prior probabilities of the hypothesis are given by
$\pi_k, k\in\untill{M-1}$, then the posterior probability after $N$
observations $y_l,l\in\until{N}$ is given by
\begin{equation*}
  p_N^k = \prob(H_k|y_1,\ldots,y_N) =
  \frac{\pi_k\Limits{l=1}{N}{\Pi}f^k_{s_l}(y_l)}
  {\Limits{j=0}{M-1}{\sum}\pi_j\left(\Limits{l=1}{N}{\Pi}f^j_{s_l}(y_l)\right)}, 
\end{equation*}

Before we state the MSPRT with switching sensors, for a given $N$, we define $\tilde{k}$ by
\[
\tilde{k}=\Limits{k\in\untill{M-1}}{}{\text{argmax }} \pi_k\Limits{l=1}{N}{\Pi} f^k_{s_l}(y_l).
\]
For the thresholds $\eta_k$, $k\in\untill{M-1}$, defined in
equation~\eqref{eq:threshold}, the MSPRT with switching sensors at each
sampling iteration $N$ is defined by
\begin{equation*}
  \begin{cases}
    p^k_n > \frac{1}{1+\eta_k}, \text{ for at least one } k ,\quad&\text{say
    }H_{\tilde{k}}, \\
    \text{otherwise,}& \mbox{continue sampling.}
  \end{cases}
\end{equation*}


Before we state the results on asymptotic sample size and expected decision
time, we introduce the following notation. For a given hypothesis $H_k$,
and a sensor $s$, we define $j^*_{(s,k)}$ by
\begin{equation*}
  j^*_{(s,k)}=\Limits{\Limits{j\ne k}{j\in\untill{M-1}}{}}{}{\text{argmin}} \dist(f_s^k,f^j_s).
\end{equation*}
We also define
$\map{\mathcal{E}_\dist}{\Delta_{n-1}\times{(\mathcal{L}^1})^n\times{(\mathcal{L}^1)}^n}{\real}$
by
\begin{equation*}
  \mathcal{E}_\dist(q,f^k,f^{j^*_k})=\Limits{s=1}{n}{\sum} q_s \dist(f_s^k,f^{j^*_{(s,k)}}_s),
\end{equation*}
where $\Delta_{n-1}$ represents the probability simplex in $\real^n$.

\begin{lemma}[Expected sample size]
  \label{lem:sample-size-msprt}
  Given thresholds $\eta_j$, $j\in\untill{M-1}$, the sample size $N$
  required for decision satisfies
  \begin{equation*}
    \frac{\Exp{N|H_k}}{-\log \eta_k} \to \frac{1}{\mathcal{E}_\dist(q,f^k,f^{j^*_k})},
  \end{equation*}
  as $\Limits{j\in\untill{M-1}}{}{\max}\eta_j\to 0$.
\end{lemma}
\begin{IEEEproof}
The proof follows from Theorem 5.1 of~\cite{CWB-VVV:94} and the observation that
\begin{equation*}
  \Limits{N\to\infty}{}{\lim} \frac{1}{N} \Limits{l=1}{N}{\sum} \log
  \frac{f^k_{s_l}(X_l)}{f^j_{s_l}(X_l)}
  = \Limits{s=1}{n}{\sum} q_s \dist(f_s^k,f^j_s).
\end{equation*}
\end{IEEEproof}
%


\begin{lemma}[Expected decision time]
  \label{decision-time-dmsprt}
  Given the processing time of the sensors $T_s, s\in\until{n}$, the
  expected decision time $T_d$ conditioned on the hypothesis $H_k$, for
  each $k\in\untill{M-1}$, is given by
  \begin{equation}\label{eq:decision-time}
    \Exp{T_d|H_k} = \frac{-\log
      \eta_k}{\mathcal{E}_\dist(q,f^k,f^{j^*_k})} \Limits{s=1}{n}{\sum}
    q_sT_s= \frac{q\cdot T}{q\cdot \bar{I}^k}, 
  \end{equation}
  where $T,\bar{I}^k\in\real^n_{>0}$ are constants.
\end{lemma}
\begin{IEEEproof}
  Similar to the proof of Lemma~\ref{lem:decision-time-switch-sprt}.
\end{IEEEproof}

\section{OPTIMAL SENSOR SELECTION}
\label{sec:optimal-sensor-selection}
In this section we consider sensor selection problems with the aim to
minimize the expected decision time of a sequential hypothesis test with
switching sensors.  As exemplified in Lemma~\ref{decision-time-dmsprt}, the
problem features multiple conditioned decision times and, therefore,
multiple distinct cost functions are of interest. In Scenario I below, we
aim to minimize the decision time conditioned upon one specific hypothesis
being true; in Scenarios II and III we will consider worst-case and average
decision times. In all three scenarios the decision variables take values
in the probability simplex.

Minimizing decision time conditioned upon a specific hypothesis may be of
interest when fast reaction is required in response to the specific
hypothesis being indeed true.  For example, in change detection problems
one aims to quickly detect a change in a stochastic process; the CUSUM
algorithm (also referred to as Page's test)~\cite{ESP:1954} is widely used
in such problems. It is known~\cite{MB-IVN:93} that, with fixed threshold,
the CUSUM algorithm for quickest change detection is equivalent to an SPRT
on the observations taken after the change has occurred. We consider the
minimization problem for a single conditioned decision time in Scenario I
below and we show that, in this case, observing the best sensor each time
is the optimal strategy.

In general, no specific hypothesis might play a special role in the problem
and, therefore, it is of interest to simultaneously minimize multiple
decision times over the probability simplex. This is a multi-objective
optimization problem, and may have Pareto-optimal solutions. We tackle this
problem by constructing a single aggregate objective function. In the
binary hypothesis case, we construct two single aggregate objective
functions as the maximum and the average of the two conditioned decision
times. These two functions are discussed in Scenario II and Scenario III
respectively. In the multiple hypothesis setting, we consider the single
aggregate objective function constructed as the average of the conditioned
decision times. An analytical treatment of this function for $M>2$, is
difficult. We determine the optimal number of sensors to be observed, and
direct the interested reader to some iterative algorithms to solve such
optimization problems. This case is also considered under Scenario III.

Before we pose the problem of optimal sensor selection, we introduce the
following notation. We denote the probability simplex in $\mathbb{R}^{n}$
by $\Delta_{n-1}$, and the vertices of the probability simplex
$\Delta_{n-1}$ by $e_i$, $i\in\until{n}$. We refer to the line joining any
two vertices of the simplex as an \emph{edge}.  Finally, we define
$\map{g^k}{\Delta_{n-1}}{\real}$, $k\in\untill{M-1}$, by
\begin{align*}
g^k(q)=\frac{q\cdot T}{q\cdot I^k}.
\end{align*}

\subsection {Scenario I {(Optimization of conditioned decision time)}:}
We consider the case when the supervisor is trying to detect a particular
hypothesis, irrespective of the present hypothesis. The corresponding
optimization problem for a fixed $k\in\untill{M-1}$ is posed in the
following way:
\begin{equation} \label{eq:minimize_one_time}
  \begin{split}
    \minimize &\quad g^k(q)\\
    \subject  &\quad q\in\Delta_{n-1}.
  \end{split}
\end{equation}

The solution to this minimization problem is given in the following theorem.

\begin{theorem}[Optimization of conditioned decision time]\label{thm:single}
  The solution to the minimization problem \eqref{eq:minimize_one_time} is
  $q^*=e_{s^*}$, where $s^*$ is given by
  \begin{equation*}
    s^*=\Limits{s\in\until{n}}{}{\mathrm{argmin}} \frac{T_s}{I_s^k},    
  \end{equation*}
  and the minimum objective function is
  \begin{equation}
    \Exp{T_d^*|H_k} =\frac{T_{s^*}}{I_{s^*}^k}.
  \end{equation}
\end{theorem}

\begin{IEEEproof}
  We notice that objective function is a linear-fractional function. In the
  following argument, we show that the minima occurs at one of the vertices
  of the simplex.

  We first notice that the probability simplex is the convex hull of the
  vertices, i.e., any point $\tilde{q}$ in the probability simplex can be
  written as
  \begin{align*}
    \tilde{q}=\Limits{s=1}{n}\sum \alpha_s e_s, \quad  \quad
    \Limits{s=1}{n}\sum \alpha_s =1,\quad \mbox{and} \quad \alpha_s\ge 0 .
  \end{align*}
  We invoke equation \eqref{eq:lin_frac}, and observe that for some $\beta \in [0,1]$ and for any $s,r\in \until{n}$
  \begin{align}
    g^k(\beta e_s + (1-\beta) e_r) \ge \mbox{min} \{g^k(e_s),g^k(e_r)\},
  \end{align}
  which can be easily generalized to
  \begin{align}
    g^k(\tilde{q}) \ge \Limits{l\in \until{n}}{}{\mbox{min}} g^k(e_s),
  \end{align}
  for any point $\tilde{q}$ in the probability simplex
  $\Delta_{n-1}$. Hence, minima will occur at one of the vertices
  $e_{s^*}$, where $s^*$ is given by
  \begin{align*}
    s^* = \Limits{s\in\until{n}} {} {\mbox{argmin}} g^k(e_s) =
    \Limits{s\in\until{n}} {} {\mbox{argmin}} \frac{T_s}{I_s^k}.
  \end{align*}
\end{IEEEproof}

\subsection{Scenario II {(Optimization of the worst case decision time)}:}
For the binary hypothesis testing, we consider the multi-objective
optimization problem of minimizing both decision times simultaneously. We
construct single aggregate objective function by considering the maximum of
the two objective functions. This turns out to be a worst case analysis,
and the optimization problem for this case is posed in the following way:

\begin{equation} \label{eq:minimize_maximum_time}
\begin{split}
\minimize \quad & \max\left\{g^0(q), g^1(q)\right\},\\
\subject \quad &  q\in\Delta_{n-1}.
\end{split}
\end{equation}

Before we move on to the solution of above minimization problem, we state
the following results.

\begin{lemma}[Monotonicity of conditioned decision times]\label{lem:mon}
  The functions $g^k$, $k\in\untill{M-1}$ are monotone on the probability
  simplex $\Delta_{n-1}$, in the sense that given two points
  $q_1,q_2\in\Delta_{n-1}$, the function $g^k$ is monotonically
  non-increasing or monotonically non-decreasing along the line joining
  $q_1$ and $q_2$.
\end{lemma}
\begin{IEEEproof}
  Consider probability vectors $q_1,q_2 \in \Delta_{n-1}$. Any point $q$ on
  line joining $q_1$ and $q_2$ can be written as $q(t)=tq_1+(1-t)q_2$,
  $t\in{]0,1[}$. We note that $g^k(q(t))$ is
  given by:
  \begin{align*}
    g^k(q(t))=\frac{t(q_1\cdot T)+(1-t)(q_2\cdot
      T)}{t(q_1\cdot{I}^k)+(1-t)(q_2\cdot{I}^k)}.
  \end{align*}

  The derivative of $g^k$ along the line joining $q_1$ and $q_2$ is given by
  \begin{multline*}
    \frac{d}{dt} g^k(q(t)) = \left(g^k(q_1)-g^k(q_2)\right) \\
    \times \frac{(q_1\cdot{I}^k)(q_2\cdot{I}^k)}{(t(q_1\cdot{I}^k)+(1-t)(q_2\cdot{I}^k))^2}.
  \end{multline*}
  We note that the sign of the derivative of $g^k$ along the line joining
  two points $q_1,q_2$ is fixed by the choice of $q_1$ and $q_2$. Hence,
  the function $g^k$ is monotone over the line joining $q_1$ and
  $q_2$. Moreover, note that if $g^k(q_1)\ne g^k(q_2)$, then $g^k$ is
  strictly monotone. Otherwise, $g^k$ is constant over the line joining
  $q_1$ and $q_2$.
\end{IEEEproof}

\begin{lemma}[Location of min-max]\label{lem:opt_max}
  Define $\map{g}{\Delta_{n-1}}{\mathbb{R}_{\ge 0}}$ by
  $g=\mbox{max}\{g^0,g^1\}$. A minimum of $g$ lies at the intersection of
  the graphs of $g^0$ and $g^1$, or at some vertex of the probability
  simplex $\Delta_{n-1}$.
\end{lemma}
\begin{IEEEproof}
{Case 1:} The graphs of $g^0$ and $g^1$ do not intersect at any point in the simplex $\Delta_{n-1}$.

In this case, one of the functions $g^0$ and $g^1$ is an upper bound to the other function at every point in the probability simplex $\Delta_{n-1}$. Hence, $g=g^k$, for some $k\in\{0,1\}$, at every point in the probability simplex $\Delta_{n-1}$. From Theorem \ref{thm:single}, we know that the minima of $g^k$ on the probability simplex $\Delta_{n-1}$ lie at some vertex of the probability simplex $\Delta_{n-1}$.

{Case 2:} The graphs of $g^0$ and $g^1$  intersect at a set $Q$ in the probability simplex $\Delta_{n-1}$, and let $\bar{q}$ be some point in the set $Q$.

Suppose, a minimum of $g$ occurs at some point $q^* \in \text{relint}(\Delta_{n-1})$, and $q^*\notin Q$, where relint$(\cdot)$ denotes the relative interior. With out loss of generality, we can assume that $g^0(q^*) > g^1(q^*)$. Also, $g^0(\bar{q}) = g^1(\bar{q})$, and $g^0(q^*)<g^0(\bar{q})$ by assumption.

We invoke Lemma \ref{lem:mon}, and notice that $g^0$ and $g^1$ can
intersect at most once on a line. Moreover, we note that
$g^0(q^*)>g^1(q^*)$, hence, along the half-line from $\bar{q}$ through
$q^*$, $g^0>g^1$, that is, $g=g^0$. Since $g^0(q^*)<g^0(\bar{q})$, $g$ is decreasing along this half-line. Hence, $g$ should achieve its minimum at the boundary of the simplex $\Delta_{n-1}$, which contradicts that $q^*$ is in the relative interior of the simplex $\Delta_{n-1}$. In summary, if a minimum of $g$ lies in the relative interior of the probability simplex $\Delta_{n-1}$, then it lies at the intersection of the graphs of $g^0$ and $g^1$.

The same argument can be applied recursively to show that if a minimum lies at some point $q^\dag$ on the boundary, then either $g^0(q^\dag)=g^1(q^\dag)$ or the minimum lies at the vertex.
\end{IEEEproof}

In the following arguments, let $Q$ be the set of points in the simplex $\Delta_{n-1}$, where $g^0=g^1$, that is,
\begin{equation}\label{eq:inter_set}
Q=\setdef{q\in \Delta_{n-1}}{ q\cdot(I^0-I^1)=0}.
\end{equation}

 Also notice that the set $Q$ is non empty if and only if $I^0-I^1$ has at least one non-negative and one non-positive entry.  If the set $Q$ is empty, then it follows from Lemma \ref{lem:opt_max} that the solution of optimization problem in equation~\eqref{eq:minimize_maximum_time} lies at some vertex of the probability simplex $\Delta_{n-1}$. Now we consider the case when $Q$ is non empty. We assume that the sensors have been re-ordered such that the entries in $I^0-I^1$ are in ascending order. We further assume that, for $I^0-I^1$, the first $m$ entries, $m<n$, are non positive, and the remaining entries are positive.

\begin{lemma}[Intersection polytope]
  \label{lem:vertex}
  If the set $Q$ defined in equation \eqref{eq:inter_set} is non empty,
  then the polytope generated by the points in the set $Q$ has vertices
  given by:
\begin{align}
\tilde{Q}= \setdef{\tilde{q}^{sr}&}{s\in\until{m} \text{ and }   r\in\{m+1,\ldots,n\}},\notag\\
\text{where for each }  &i\in\until{n}\notag\\
 \tilde{q}^{sr}_i&=\begin{cases}
\frac{(I_r^0-I_r^1)}{(I_r^0-I_r^1)-(I_s^0-I_s^1)}, & \text{if} \quad i=s,\\
1-\tilde{q}^{sr}_s, & \text{if} \quad i=r,\\
0, & \text{otherwise}.
\end{cases}\label{eq:vertex2}
\end{align}
\end{lemma}
\begin{IEEEproof}
Any $q\in Q$ satisfies the following constraints
\begin{gather}
\Limits{s=1}{n}{\sum} q_s  =1, \quad  q_s\in[0,1], \label{eq:simplex}\\
\Limits{s=1}{n}{\sum} q_s(I^0_s-I^1_s) =0, \label{eq:null}
\end{gather}
Eliminating $q_n$, using equation \eqref{eq:simplex} and equation \eqref{eq:null}, we get:
\begin{align}\label{eq:hyper}
\Limits{s=1}{n-1}{\sum} &\beta_sq_s =1, \quad \text{where} \quad \beta_s=\frac{(I_n^0-I_n^1)-(I_s^0-I_s^1)}{(I_n^0-I_n^1)}.
\end{align}
The equation \eqref{eq:hyper} defines a hyperplane, whose extreme points in $\mathbb{R}^{n-1}_{\ge 0}$ are given by
\begin{align*}
\tilde{q}^{sn}=\frac{1}{\beta_s} e_s, \quad i\in\until{n-1}.
\end{align*}

Note that for $s\in\until{m}$, $\tilde{q}^{sn}\in\Delta_{n-1}$. Hence, these points define some vertices of the polytope generated by points in the set $Q$. Also note that the other vertices of the polytope can be determined by the intersection of each pair of lines through $\tilde{q}^{sn}$ and $\tilde{q}^{rn}$, and $e_s$ and $e_r$, for $s\in\until{m}$, and $r\in\{m+1,\ldots,n-1\}$. In particular, these vertices are given by $\tilde{q}^{sr}$ defined in equation \eqref{eq:vertex2}.

Hence, all the vertices of the polytopes are defined by $\tilde{q}^{sr}$, $s\in\until{m}$, $r\in\{m+1,\ldots,n\}$.  Therefore, the set of vertices of the polygon generated by the points in the set $Q$ is $\tilde{Q}$.
\end{IEEEproof}

Before we state the solution to the optimization problem \eqref{eq:minimize_maximum_time}, we define the following:
\begin{align*}
&(s^*,r^*)\in\Limits{\Limits{s\in\until{m}}{r\in\{m+1,\ldots,n\}}{}{}}{}{\text{argmin}}
\frac{(I_r^0-I_r^1)T_s-(I_s^0-I_s^1)T_r}{I_s^1I_r^0-I_s^0I_r^1},\quad\text{and}\\
&g_{\textup{two-sensors}}(s^*,r^*)=\frac{(I_{r^*}^0-I_{r^*}^1)T_{s^*}-(I_{s^*}^0-I_{s^*}^1)T_{r^*}}{I_{s^*}^1I_{r^*}^0-I_{s^*}^0I_{r^*}^1}.
\end{align*}
We also define
\begin{align*}
  w^*&=\Limits{w\in\until{n}}{}{\text{argmin}} \text{max}
  \left\{\frac{T_w}{I_w^0},\frac{T_w}{I_w^1}\right\},\quad\text{and} 
  \\
  g_{\text{one-sensor}}(w^*)&=\text{max}\left\{\frac{T_{w^*}}{I^0_{w^*}},\frac{T_{w^*}}{I^1_{w^*}}\right\}.
\end{align*}
\begin{theorem}[Worst case optimization]\label{thm:worst-case}
 For the optimization problem \eqref{eq:minimize_maximum_time}, an optimal probability vector is given by:
\begin{align*}
q^*=
\begin{cases}
e_{w^*}, & \text{if } g_{\text{one-sensor}}(w^*) \le g_{\textup{two-sensors}}(s^*,r^*),\\
\tilde{q}^{s^*r^*}, & \text{if } g_{\text{one-sensor}}(w^*) > g_{\textup{two-sensors}}(s^*,r^*),
\end{cases}
\end{align*}
and the minimum value of the function is given by:
\begin{align*}
\text{min}\left\{g_{\text{one-sensor}}(w^*),g_{\textup{two-sensors}}(s^*,r^*)\right\}.
\end{align*}
\end{theorem}

\begin{IEEEproof}
  We invoke Lemma \ref{lem:opt_max}, and note that a minimum should lie at
  some vertex of the simplex $\Delta_{n-1}$, or at some point in the set
  $Q$. Note that $g^0=g^1$ on the set $Q$, hence the problem of minimizing
  max$\{g^0,g^1\}$ reduces to minimizing $g^0$ on the set $Q$. From Theorem
  \ref{thm:single}, we know that $g^0$ achieves the minima at some extreme
  point of the feasible region. From Lemma \ref{lem:vertex}, we know that
  the vertices of the polytope generated by points in set $Q$ are given by
  set $\tilde{Q}$. We further note that $g_{\textup{two-sensors}}(s,r)$ and
  $g_{\text{one-sensor}}(w)$ are the value of objective function at the
  points in the set $\tilde{Q}$ and the vertices of the probability simplex
  $\Delta_{n-1}$ respectively, which completes the proof.
\end{IEEEproof}

\subsection{Scenario III {(Optimization of the average decision time)}:} 
For the multi-objective optimization problem of minimizing all the decision
times simultaneously on the simplex, we formulate the single aggregate
objective function as the average of these decision times. The resulting
optimization problem, for $M\ge 2$, is posed in the following way:
\begin{equation} \label{eq:min_sum}
\begin{split}
\minimize\quad &\frac{1}{M} (g^0(q)+\ldots+g^{M-1}(q)),\\
\subject \quad & q\in\Delta_{n-1}.
\end{split}
\end{equation}

In the following discussion we assume $n>M$, unless otherwise stated. We
analyze the optimization problem in equation~\eqref{eq:min_sum} as follows:
\begin{lemma}[Non-vanishing Jacobian] 
  \label{lem:monotone-sum} 
  The objective function in optimization problem in
  equation~\eqref{eq:min_sum} has no critical point on $\Delta_{n-1}$ if
  the vectors $T,I^0,\ldots,I^{M-1} \in \mathbb{R}^n_{>0}$ are linearly
  independent.
\end{lemma}
\begin{IEEEproof}
The Jacobian of the objective function in the optimization problem in equation~\eqref{eq:min_sum} is
\begin{align*}
\frac{1}{M}\frac{\partial}{\partial q}&\Limits{k=0}{M-1}{\sum}g^k =\Gamma \psi(q),\\
\text{where }
\Gamma &=
\frac{1}{M}\left[\begin{array}{cccc}
    \!\!T & -I^0 & \ldots & -I^{M-1}\!\!\!
  \end{array}\right]\in\real^{n\times (M+1)}, \text{ and} \\
\map{\psi}{\Delta_{n-1}&}{\real^{M+1}} \text{ is defined by }\\
\psi(q) &=
 \left[\begin{array}
{cccc} \Limits{k=0}{M-1}{\sum} \frac{1}{q\cdot I^k}&
\frac{q\cdot T}{(q\cdot I^0)^2} &
\ldots&
\frac{q\cdot T}{(q\cdot I^{M-1})^2}
\end{array}\right]^{\text{T}}.
\end{align*}

For $n>M$, if the vectors $T,I^0,\ldots,I^{M-1}$ are linearly independent, then $\Gamma$ is full rank. Further, the entries of $\psi$ are non-zero on the probability simplex $\Delta_{n-1}$. Hence, the Jacobian does not vanish anywhere on the probability simplex $\Delta_{n-1}$.
\end{IEEEproof}

\begin{lemma}[Case of Independent Information]\label{lem:indep}
  For $M=2$, if $I^0$ and $I^1$ are linearly independent, and $T=\alpha_0
  I^0+\alpha_1 I^1$, for some $\alpha_0, \alpha_1 \in \mathbb{R}$, then the
  following statements hold:
  \begin{enumerate}
  \item if $\alpha_0$ and $\alpha_1$ have opposite signs, then $g^0+g^1$
    has no critical point on the simplex $\Delta_{n-1}$, and
  \item for $\alpha_0, \alpha_1 > 0$, $g^0+g^1$ has a critical point on the
    simplex $\Delta_{n-1}$ if and only if there exists $v \in \Delta_{n-1}$
    perpendicular to the vector $\sqrt{\alpha_0}I^0-\sqrt{\alpha_1} I^1$.
  \end{enumerate}
\end{lemma}
\begin{IEEEproof}
  We notice that the Jacobian of $g^0+g^1$ satisfies
  \begin{equation}
    \label{eq:ind-info}
    \begin{split}
    (q\cdot I^0)^2&(q\cdot I^1)^2\frac{\partial}{\partial q}(g^0+g^1)
    \\
    =&\; T\left((q\cdot I^0) (q\cdot I^1)^2+ (q\cdot I^1) (q\cdot I^0)^2\right)\\
    &\; -I^0(q\cdot T)(q\cdot I^1)^2 -I^1(q\cdot T)(q\cdot I^0)^2.      
    \end{split}
  \end{equation}
  Substituting $T=\alpha_0 I^0 + \alpha_1 I^1$, equation~\eqref{eq:ind-info}  becomes
  \begin{multline*}
    (q\cdot I^0)^2(q\cdot I^1)^2\frac{\partial}{\partial q}(g^0+g^1)\\
    =\left(\alpha_0 (q \cdot I^0)^2 - \alpha_1 (q \cdot I^1)^2 \right)\left((q\cdot I^1)I^0 - (q\cdot I^0)I^1\right).
  \end{multline*}
  Since $I^0$, and $I^1$ are linearly independent, we have
  \begin{align*}
    &\frac{\partial}{\partial q}(g^0+g^1)=0 \iff \alpha_0 (q \cdot I^0)^2 - \alpha_1 (q \cdot I^1)^2 =0.
  \end{align*}
  Hence, $g^0+g^1$ has a critical point on the simplex $\Delta_{n-1}$ if and only if
  \begin{align} \label{eq:cond} \alpha_0 (q \cdot I^0)^2 &= \alpha_1 (q \cdot
    I^1)^2.
  \end{align}
  Notice that, if $\alpha_0$, and $\alpha_1$ have opposite signs, then
  equation \eqref{eq:cond} can not be satisfied for any $q\in\Delta_{n-1}$,
  and hence, $g^0+g^1$ has no critical point on the simplex $\Delta_{n-1}$.
  
  If $\alpha_0, \alpha_1 >0$, then equation \eqref{eq:cond} leads to
  \begin{align*}
    q\cdot (\sqrt{\alpha_0} I^0- \sqrt{\alpha_1} I^1) =0.
  \end{align*}
  Therefore, $g^0+g^1$ has a critical point on the simplex $\Delta_{n-1}$
  if and only if there exists $v \in \Delta_{n-1}$ perpendicular to the
  vector $\sqrt{\alpha_0} I^0- \sqrt{\alpha_1} I^1$.
\end{IEEEproof}

\begin{lemma}[Optimal number of sensors] 
  \label{lem:optimal-sensors} 
  For $n>M$, if each $(M+1)\times(M+1)$ submatrix of the matrix
  \[
  \Gamma= \left[\begin{array}{cccc}
      T & -I^0 & \ldots & -I^{M-1}
    \end{array}\right]\in\real^{n\times (M+1)}
  \]
  is full rank, then the following statements hold:
  \begin{enumerate}
  \item every solution of the optimization problem~\eqref{eq:min_sum} lies
    on the probability simplex $\Delta_{M-1}\subset\Delta_{n-1}$; and
  \item every time-optimal policy requires at most $M$ sensors to be
    observed.
  \end{enumerate}
\end{lemma}
\begin{IEEEproof}
  From Lemma \ref{lem:monotone-sum}, we know that if $T,
  I^0,\ldots,I^{M-1}$ are linearly independent, then the Jacobian of the
  objective function in equation~\eqref{eq:min_sum} does not vanish
  anywhere on the simplex $\Delta_{n-1}$. Hence, a minimum lies at some
  simplex $\Delta_{n-2}$, which is the boundary of the simplex
  $\Delta_{n-1}$. Notice that, if $n>M$ and the condition in the lemma
  holds, then the projections of $T,I^0,\ldots,I^{M-1}$ on the simplex
  $\Delta_{n-2}$ are also linearly independent, and the argument
  repeats. Hence, a minimum lies at some simplex $\Delta_{M-1}$, which
  implies that optimal policy requires at most $M$ sensors to be observed.
\end{IEEEproof}

\begin{lemma}[Optimization on an edge]\label{lem:opt-edge}
  Given two vertices $e_s$ and $e_r$, $s\ne r$, of the probability simplex
  $\Delta_{n-1}$, then for the objective function in the problem
  \eqref{eq:min_sum} with $M=2$, the following statements hold:
  \begin{enumerate}
  \item if $g^0(e_s)<g^0(e_r)$, and $g^1(e_s)<g^1(e_r)$, then the minima,
    along the edge joining $e_s$ and $e_r$, lies at $e_s$, and optimal
    value is given by $\frac{1}{2}(g^0(e_s)+g^1(e_s))$; and
  \item if $g^0(e_s)>g^0(e_r)$, and $g^1(e_s)<g^1(e_r)$, or vice versa,
    then the minima, along the edge joining $e_s$ and $e_r$, lies at the
    point $q^*=(1-t^*)e_s+t^*e_r$, where
    \begin{align*}
      t^*&=\frac{1}{1+\mu} \in {]0,1[},\\
      \mu&=\frac{I_r^0\sqrt{T_s I_r^1 -T_r
          I_s^1}-I_r^1\sqrt{T_rI_s^0-T_sI_r^0}}{I_s^1\sqrt{T_rI_s^0-T_sI_r^0}-I^0_s\sqrt{T_s
          I_r^1 -T_r I_s^1}}>0,
    \end{align*}
    and the optimal value is given by
    \begin{multline*}
      \frac{1}{2}(g^0(q^*)+g^1(q^*)) \\
      = \frac{1}{2}\left( {\sqrt{\frac{T_s I_r^1 -T_r
              I_s^1}{I_s^0I_r^1-I_r^0I_s^1}}+
          \sqrt{\frac{T_rI_s^0-T_sI_r^0}{I_s^0I_r^1-I_r^0I_s^1}}}
      \right)^2.
    \end{multline*}
\end{enumerate}
\end{lemma}

\begin{IEEEproof}
  We observe from Lemma \ref{lem:mon} that both $g^0$, and $g^1$ are
  monotonically non-increasing or non-decreasing along any line. Hence, if
  $g^0(e_s)<g^0(e_r)$, and $g^1(e_s)<g^1(e_r)$, then the minima should lie
  at $e_s$. This concludes the proof of the first statement.  We now
  establish the second statement. We note that any point on the line
  segment connecting $e_s$ and $e_r$ can be written as
  $q(t)=(1-t)e_s+te_r$. The value of $g^0+g^1$ at $q$ is
\begin{align*}
g^0(q(t))+g^1(q(t))= \frac{(1-t)T_s+tT_r}{(1-t)I_s^0+tI_r^0} + \frac{(1-t)T_s+tT_r}{(1-t)I_s^1+tI_r^1} .
\end{align*}
Differentiating with respect to $t$, we get
\begin{multline}\label{eq:der}
{g^0}'(q(t))+{g^1}'(q(t))\\
=\frac{I^0_sT_r-T_sI^0_r}{(I_s^0+t(I_r^0-I_s^0))^2} + \frac{I^1_sT_r-T_sI^1_r}{(I_s^1+t(I_r^1-I_s^1))^2} .
\end{multline}

Notice that the two terms in equation \eqref{eq:der} have opposite sign. Setting the derivative to zero, and choosing the value of $t$ in $[0,1]$, we get $t^*=\frac{1}{1+\mu}$, where $\mu$ is as defined in the statement of the theorem.
The optimal value of the function can be obtained, by substituting $t=t^*$ in the expression for $\frac{1}{2}(g^0(q(t))+g^1(q(t)))$.
\end{IEEEproof}

\begin{theorem}[Optimization of average decision time]\label{thm:average}
For the optimization problem in equation~\eqref{eq:min_sum} with $M=2$, the following statements hold:
\begin{enumerate}
\item If $I^0$, $I^1$ are linearly dependent, then the solution lies at some vertex of the simplex $\Delta_{n-1}$.
\item If $I^0$ and $I^1$ are linearly independent, and $T=\alpha_0 I^0 + \alpha_1 I^1$, $\alpha_0, \alpha_1 \in \mathbb{R}$, then the following statements hold:
\begin{enumerate}
\item If $\alpha_0$ and $\alpha_1$ have opposite signs, then the optimal solution lies at some edge of the simplex $\Delta_{n-1}$.
\item If $\alpha_0, \alpha_1 >0$, then the optimal solution may lie in the interior of the simplex $\Delta_{n-1}$.
\end{enumerate}
\item If every $3\times 3$ sub-matrix of the matrix $\left[T\quad I^0 \quad I^1\right]\in\real^{n\times 3}$ is full rank, then a minimum lies at an edge of the simplex $\Delta_{n-1}$.
\end{enumerate}
\end{theorem}
\begin{IEEEproof}
We start by establishing the first statement. Since, $I^0$ and $I^1$ are linearly dependent, there exists a $\gamma>0$ such that $I^0=\gamma I^1$. For $I^0=\gamma I^1$, we have $g^0+g^1=(1+\gamma)g^0$. Hence, the minima of $g^0+g^1$ lies at the same point where $g^0$ achieves the minima. From Theorem~\ref{thm:single}, it follows that $g^0$ achieves the minima at some vertex of the simplex $\Delta_{n-1}$.

To prove the second statement, we note that from Lemma~\ref{lem:indep}, it follows that if $\alpha_0$, and $\alpha_1$ have opposite signs, then the Jacobian of $g^0+g^1$ does not vanish anywhere on the simplex $\Delta_{n-1}$. Hence, the minima lies at the boundary of the simplex. Notice that the boundary, of the simplex $\Delta_{n-1}$, are $n$ simplices $\Delta_{n-2}$. Notice that the argument repeats till $n>2$. Hence, the optima lie on one of the $\binom{n}{2}$ simplices $\Delta_1$, which are the edges of the original simplex. Moreover, we note that from Lemma~\ref{lem:indep}, it follows that if $\alpha_0, \alpha_1>0$, then we can not guarantee the number of optimal set of sensors. This concludes the proof of the second statement.

To prove the last statement, we note that it follows immediately from Lemma~\ref{lem:optimal-sensors} that a solution of the optimization problem in equation~\eqref{eq:min_sum} would lie at some simplex $\Delta_1$, which is an edge of the original simplex.
\end{IEEEproof}

Note that, we have shown that, for $M=2$ and a generic set of sensors, the solution of the optimization problem in equation~\eqref{eq:min_sum} lies at an edge of the simplex $\Delta_{n-1}$. The optimal value of the objective function on a given edge was determined in Lemma~\ref{lem:opt-edge}. Hence, an optimal solution of this problem can be determined by a comparison of the optimal values at each edge.


For the multiple hypothesis case, we have determined the time-optimal
number of the sensors to be observed in Lemma~\ref{lem:optimal-sensors}. In
order to identify these sensors, one needs to solve the optimization
problem in equation~\eqref{eq:min_sum}. We notice that the objective
function in this optimization problem is non-convex, and is hard to tackle
analytically for $M>2$. Interested reader may refer to some efficient
iterative algorithms in linear-fractional programming literature (e.g., \cite{HPB:04}) to solve these problems. 


\section{NUMERICAL EXAMPLES}\label{sec:numerical-examples}

We consider four sensors connected to a fusion center. We assume that the sensors take binary measurements. The probabilities of their measurement being zero, under two hypotheses, and their processing times are given in the Table~\ref{tab:prob}.

\begin{table}[ht]
\caption{Conditional probabilities of measurement being zero}
\begin{center}
\begin{tabular}{ c | c |c|c }\label{tab:prob}
Sensor{}& \multicolumn{2}{c|} {Probability(0)}&Processing Time\\
\hline
{}& Hypothesis 0 & Hypothesis 1 &{} \\
\hline
1 & 0.4076 & 0.5313 & 0.6881\\
\hline
2 & 0.8200 & 0.3251 & 3.1960 \\
\hline
3 & 0.7184 & 0.1056 & 5.3086\\
\hline
4 & 0.9686 & 0.6110 & 6.5445
\end{tabular}
\end{center}
\end{table}

We performed Monte-Carlo simulations with the mis-detection and false-alarm probabilities fixed at $10^{-3}$, and computed expected decision times. In Table~\ref{tab:dec_time}, the numerical expected decision times are compared with the decision times obtained analytically in equation~\eqref{eq:expected_decision_time}. The difference in the numerical and the analytical decision times is explained by the Wald's asymptotic approximations.

\begin{table}[ht]
\caption{Decision times for various sensor selection probabilities}
\begin{center}
\resizebox{\linewidth}{!}{
\begin{tabular}{ c | c | c |c |c }\label{tab:dec_time}
Sensor selection &\multicolumn{4}{c}{Expected Decision Time}   \\
probability &\multicolumn{4}{c}{}\\
\hline
{} &\multicolumn{2}{c|}{Hypothesis 0}  & \multicolumn{2}{c}{Hypothesis 1}  \\
\hline
{} & Analytical & Numerical & Analytical & Numerical\\
\hline
[1,0,0,0] & 154.39 & 156.96 & 152.82 & 157.26\\
\hline
[0.3768,0,0,0.6232] & 124.35 & 129.36 & 66.99 & 76.24\\
\hline
[0.25,0.25,0.25,0.25] & 55.04 & 59.62 & 50.43 & 55.73
\end{tabular}}
\end{center}
\end{table}
\normalsize
In the Table~\ref{tab:opt_comp}, we compare optimal policies in each Scenario I, II, and III with the policy when each sensor is chosen uniformly. It is observed that the optimal policy improves the expected decision time significantly over the uniform policy.
\begin{table}[ht]
\caption{Comparison of the uniform and optimal policies}
\begin{center}
\begin{tabular}{ c | c |c |c }\label{tab:opt_comp}
Scenario & Uniform policy  & \multicolumn{2}{c}{Optimal policy}  \\
\hline
{}  &  Objective & Optimal probability & Optimal objective \\
{}& function & vector & function\\
\hline
I & 59.62 & [0,0,1,0] & 41.48 \\
\hline
II &  55.73 &[0,0,1,0]  & 45.24  \\
\hline
III & 57.68 & [0,0.4788,0.5212,0]& 43.22
\end{tabular}
\end{center}
\end{table}

We performed another set of simulations for the multi-hypothesis case. We considered a ternary detection problem, where the underlying signal $x=0,1,2$ needs to be detected from the available noisy data. We considered a set of four sensors and their conditional probability distribution is given in Tables~\ref{tab:prob-ternary1} and~\ref{tab:prob-ternary2}. The processing time of the sensors were chosen to be the same as in Table~\ref{tab:prob}.

\begin{table}[ht]
\caption{Conditional probabilities of measurement being zero}
\begin{center}
\begin{tabular}{ c | c |c|c }\label{tab:prob-ternary1}
Sensor{}& \multicolumn{3}{c} {Probability(0)}\\
\hline
{}& Hypothesis 0 & Hypothesis 1 & Hypothesis 2 \\
\hline
1 & 0.4218  & 0.2106  & 0.2769\\
\hline
2 & 0.9157  & 0.0415  & 0.3025  \\
\hline
3 & 0.7922 & 0.1814   & 0.0971 \\
\hline
4 & 0.9595 & 0.0193 & 0.0061
\end{tabular}
\end{center}
\end{table}

\begin{table}[ht]
\caption{Conditional probabilities of measurement being one}
\begin{center}
\begin{tabular}{ c | c |c|c }\label{tab:prob-ternary2}
Sensor{}& \multicolumn{3}{c} {Probability(1)}\\
\hline
{}& Hypothesis 0 & Hypothesis 1 & Hypothesis 2 \\
\hline
1 & 0.1991   & 0.6787   & 0.2207 \\
\hline
2 & 0.0813   & 0.7577   & 0.0462  \\
\hline
3 & 0.0313 & 0.7431    & 0.0449  \\
\hline
4 & 0.0027 & 0.5884 &0.1705
\end{tabular}
\end{center}
\end{table}
The set of optimal sensors were determined for this set of
data. Monte-Carlo simulations were performed with the thresholds $\eta_k,
k\in\untill{M-1}$ set at $10^{-6}$. A comparison of the uniform sensor
selection policy and an optimal sensor selection policy is presented in
Table~\ref{tab:comparision}. Again, the significant difference between the
average decision time in the uniform and the optimal policy is evident.

\begin{table}[ht]
\caption{Comparisons of the uniform and an optimal policy }
\begin{center}
\begin{tabular}{ c |c| c  |c }\label{tab:comparision}
{Policy}& Selection Probability &\multicolumn{2}{c}{Average Decision Time}   \\
\hline
{} & $q^*$&Analytical & Numerical\\
\hline
Optimal &[0, 0.9876,  0,  0.0124] & 38.57 & 42.76 \\
\hline
Uniform&[0.25, 0.25, 0.25, 0.25] & 54.72 & 54.14 \\
\end{tabular}
\end{center}
\end{table}

We note that the optimal results, we obtained, may only be sub-optimal
because of the asymptotic approximations in
equations~\eqref{eq:thresholds}~and~\eqref{eq:threshold}. We further note
that, for small error probabilities and large sample sizes, these
asymptotic approximations yield fairly accurate results~\cite{CWB-VVV:94},
and in fact, this is the regime in which it is of interest to minimize the
expected decision time. Therefore, for all practical purposes the obtained
optimal scheme is very close to the actual optimal scheme.

\section{CONCLUSIONS}
\label{sec:conclusions}
In this paper, we considered the problem of sequential decision making. We
developed versions SPRT and MSPRT where the sensor switches at each
observation. We used these sequential procedures to decide reliably. We
found out the set of optimal sensors to decide in minimum time.  For the
binary hypothesis case, three performance metrics were considered and it
was found that for a generic set of sensors at most two sensors are
optimal. Further, it was shown that for $M$ underlying hypotheses, and a
generic set of sensors, an optimal policy requires at most $M$ sensors to
be observed. A procedure for identification of the optimal sensor was
developed. In the binary hypothesis case, the computational complexity of
the procedure for the three scenarios, namely, the conditioned decision
time, the worst case decision time, and the average decision time, was
$\mathcal{O}(n)$, $\mathcal{O}(n^2)$, and $\mathcal{O}(n^2)$, respectively.

Many further extensions to the results presented here are possible.  First,
the time-optimal scheme may not be energy optimal. In particular, the time
optimal set of sensors may be the most distant sensors from the fusion
center. Given that the power to transmit the signal to the fusion center is
proportional to the distance from the fusion center, the time-optimal
scheme is no where close to the energy optimal scheme.  This trade off can
be taken care of by adding a term proportional to distance in the objective
function.

When we choose only one or two sensors every time, issues of robustness do
arise. In case of sensor failure, we need to determine the next best sensor
to switch.  A list of sensors, with increasing decision time, could be
prepared beforehand and in case of sensor failure, the fusion center should
switch to the next best set of sensors.



\end{document}